\documentclass[conference]{IEEEtran}
\ifCLASSINFOpdf
\else
\fi
\pdfoutput=1
\usepackage{booktabs} 
\usepackage{caption}
\usepackage{subcaption}
\usepackage{amsthm}
\usepackage{amsmath,amssymb}
\usepackage{mathtools}
\usepackage[bookmarks=false]{hyperref}
\usepackage{algorithm}
\usepackage[noend]{algpseudocode}
\usepackage{color}
\usepackage{colortbl}
\usepackage{amssymb}
 
\newcommand*{\affaddr}[1]{#1} 
\newcommand*{\affmark}[1][*]{\textsuperscript{#1}}
\newcommand*{\email}[1]{\texttt{#1}}
\usepackage[pdftex]{graphicx}

\begin{document}
%

\title{Core2Vec: A core-preserving feature\\ learning framework for networks}
\small

\author{%
Soumya Sarkar\affmark[1], Aditya Bhagwat\affmark[2], Animesh Mukherjee\affmark[3]\\
\affaddr{Department of Computer Science and Engineering, IIT Kharagpur, India}\\
\email{soumya015@iitkgp.ac.in\affmark[1],bhagwat.work@gmail.com\affmark[2]}\\
\email{animesh@cse.iitkgp.ernet.in\affmark[3]}\\
}


%


\maketitle

\begin{abstract}
Recent advances in the field of network representation learning are mostly attributed to the application of the skip-gram model in the context of graphs. State-of-the-art analogues of skip-gram model in graphs define a notion of neighbourhood and aim to find the vector representation for a node, which maximizes the likelihood of preserving this neighborhood.

In this paper, we take a drastic departure from the existing notion of neighbourhood of a node by utilizing the idea of \textit{coreness}. More specifically, we utilize the well-established idea that nodes with similar core numbers play equivalent roles in the network and hence induce a novel and an organic notion of neighbourhood. Based on this idea, we propose \textit{core2vec}, a new algorithmic framework for learning low dimensional continuous feature mapping for a node. Consequently, the nodes having similar core numbers are relatively closer in the vector space that we learn.

We further demonstrate the effectiveness of \textit{core2vec} by comparing word similarity scores obtained by our method where the node representations are drawn from standard word association graphs\footnote{In linguistics, such networks built from various linguistic units are known to have a core-periphery structure (see~\cite{choudhury2010global} and the references therein).} against scores computed by other state-of-the-art network representation techniques like node2vec, DeepWalk and LINE. Our results always outperform these existing methods, in some cases achieving improvements as high as \textbf{46\%} on certain ground-truth word similarity datasets. We make all codes used in this paper available in the public domain:  \url{https://github.com/Sam131112/Core2vec_test}.

\end{abstract}

\IEEEpeerreviewmaketitle
\section{Introduction}

The conventional paradigm of handcrafted feature engineering to generate node representations in networks has been largely overhauled due to advances in techniques which automatically discover and map a node's structural properties into a latent space. These techniques are useful because manual feature engineering requires extensive domain knowledge as well as tedious exploration of structural properties such as degree, centrality, clustering coefficient etc. Without loss of generality, representation learning encompasses the task of  transforming a graph $G(V,E)$ from $V \rightarrow \mathbb{I}^{|V| \times |V| }$ to the mapping $V \rightarrow \mathbb{R}^{|V| \times d }$ with the constraint $d <<|V|$.

\subsection{Random walk based techniques}
This problem is efficiently solved by applying a {\em skip-gram model with negative sampling}\cite{mikolov2013distributed}, which is a celebrated technique for learning meaningful vector representations for words. To represent a target word, nearby co-occurring words in the sentence are considered as context words. Adapting this framework for graphs, there have been several works such as~\cite{perozzi2014deepwalk}\cite{tang2015line} which learn social representations of a graph's vertices, by learning from neighbor nodes generated from short random walks. {\em These walk sequences act as proxy for context words in a sentence}. Apart from capturing local proximity, global information is also captured~\cite{grover2016node2vec} through generation of flexible contexts by parameterized random walks.

\subsection{Limitations of random walk based techniques}
One of the key drawbacks in these works is the assumption that the context nodes can be always efficiently generated by walk sequences from a source node thus building a sample set appropriately representing the structural and the functional properties of the source node. This perhaps is a fair assumption in social networks which are inherently assortative. However, this  might not be applicable for several classes of networks such as biological (protein interaction), technological (router-router interaction), and semantic (e.g., Wordnet) networks which are disassortative.

\subsection{Our proposal}
We propose a solution (see section~\ref{sec:method}) to the above problem by developing an algorithmic framework, {\em core2vec} which utilizes intermediate-scale structure of the network, i.e., the core periphery structure, for learning feature representation of a node. A core-periphery structure in its simplest form refers to a partition of a network into two groups of nodes called core and periphery, where core nodes are densely interconnected (i.e., adjacent), and peripheral nodes are adjacent to the core nodes but not to other peripheral nodes. Myriad techniques exist~\cite{batagelj2003m},~\cite{rombach2017core} which attempt to discover multiple nested cores in the network. This partition of the network into nested cores of disjoint layers represent separate structural/functional properties of nodes in the network.

\begin{figure}
  \centering
  \begin{minipage}[!t]{\textwidth}
  	\includegraphics[width=0.5\linewidth, height=0.12\textheight]{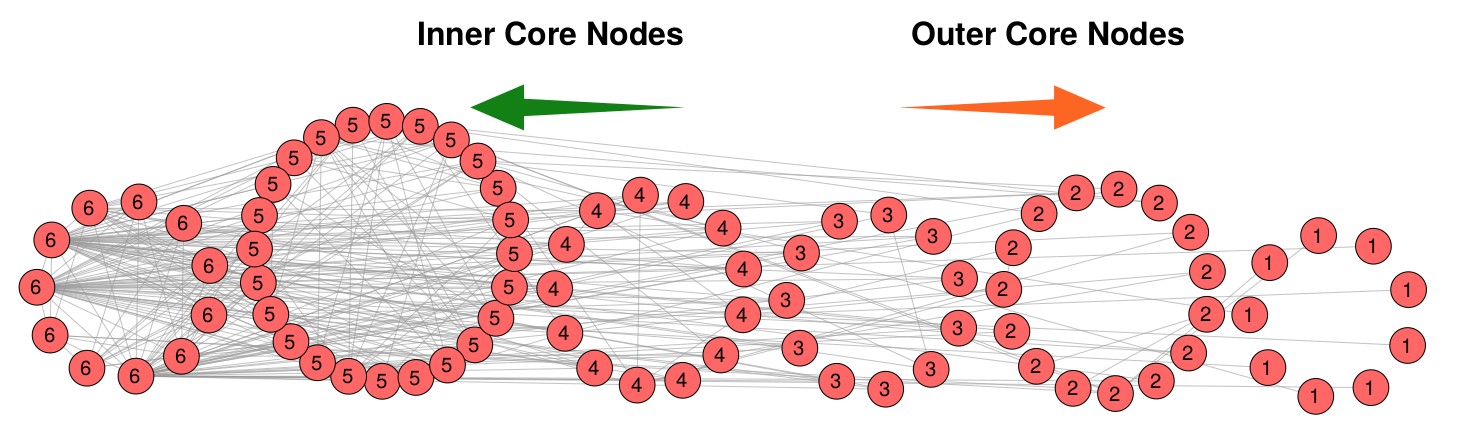}
   \end{minipage}
   \caption{Toy example showing connectivity profiles in each core. Nodes in similar core play similar roles as is evident from the figure though they may be non-adjacent. \label{connects_1}}
\end{figure}

We leverage this nested ``onion like structure" in real world complex networks, to develop a flexible biased random walk which seeks similar core nodes as context nodes for a source vertex. More specifically we develop a strategy to guide a random walk sequence to identify similar core nodes both in close proximity as well as distant neighborhood. We further design an objective function, which computes the average likelihood of predicting the source node given the set of context nodes, which we obtain through our exploration procedure. This objective function can be optimized efficiently using stochastic gradient descent and consequently leads us to our optimal set of feature vectors. Core information for a node can be computed efficiently in $O(|E|)$~\cite{batagelj2003m} and nodes with similar core ids have equivalent connectivity profiles (see Fig~\ref{connects_1}) over the entire network. We perform experiments with real world networks to analyze the performance of our scheme (see section~\ref{sec:experiments}). These experiments show that our scheme brings nodes with similar core ids closer (\textit{closeness}) as well as separates nodes with different core ids (\textit{separability}) farther in the vector space compared to state-of-the-art methods like node2vec~\cite{grover2016node2vec}, DeepWalk~\cite{perozzi2014deepwalk} and LINE~\cite{tang2015line}, thus establishing the necessity of our approach.

\subsection{Validation of core2vec}
We validate the effectiveness of our scheme by estimating similarity of words (nodes) in word association networks (see section~\ref{sec:eval}). Networks built from linguistic units (e.g., word co-occurrence and word association networks) are known to have a well-defined core-periphery structure (see~\cite{choudhury2010global} and the references therein) and hence the motivation to choose word association networks for our validation purpose.

We learn representations of each node in two large word association networks (see Table~\ref{Tab:large} and~\ref{Tab:small}) using core2vec as well as using state-of-the-art network representation learning techniques like node2vec~\cite{grover2016node2vec}, DeepWalk~\cite{perozzi2014deepwalk} and LINE~\cite{tang2015line}. We next estimate the (cosine) similarity of the vector representations of word pairs, rank these word pairs based on the similarity obtained and compare the same with the ranking of the pairs drawn from different ground-truth datasets on word similarity. We compare the rankings using the Spearman's rank correlation co-efficient and show that we always outperform the baselines. In some cases, we achieve improvements as high as \textbf{46\%} on certain ground-truth datasets.    

\section{Preamble}
\subsection{Core periphery structure}
\label{theory}
Consider an undirected graph $G(V,E)$. Let $ K \subseteq V $ and $G(K)$ be the graph induced on $G$ by the vertices in $K$. $G(K)$ is considered to be $k$-core of the graph $G$ if and only if 

(i) For every $v \in K$, $d_{G(K)}(v) \geq k $ where $d_{G(K)}(v)$ denotes the degree of $v$ in $G(K)$;

(ii) For each $K \subset K' \subset V $ $\exists$ $u \in K' \backslash K $ such that $d_{G(K')}(u) < k $.\cite{rombach2017core}.



\subsection{Objective function}
We exploit the idea of a skip-gram model with negative sampling, which is popular for language modeling, in the context of graphs. Given a set of vertices $\mathbb{C}_v$ our task is to maximize $P(\mathbb{C}_v|v)$ where $\mathbb{C}_v$ comprises the context vertices of $v$.

To complete our objective we formulate the optimization problem as 
\begin{equation}
\small
\underset{max}{f} = \sum_{v \in V} P (\mathbb{C}_v | v)  = \sum_{v \in V}  \prod_{\mathbb{C}_{v_{i}} \in \mathbb{C}_v} P( \mathbb{C}_{v_{i}}| v)
\end{equation}

where $\mathbb{C}_{v_{i}}$ is a context node belonging to the context set of $v$. $P( \mathbb{C}_{v_{i}}|v)$ can be computed as 
\begin{equation}\label{Eq:2}
\small
P(\mathbb{C}_{v_{i}}|v) = \frac{\exp(v^{w}.\mathbb{C}^{w}_{v_{i}})}{\sum_{v \in V}\exp(v^{w}.\mathbb{C}^{w}_{v_{i}})}
\end{equation}
Here $v^{w},\mathbb{C}^{w}_{v_{i}}$ denotes the vector representation of the node $v$ and context node $\mathbb{C}_{v_{i}}$. Since this formulation is difficult to optimize directly we introduce negative sampling analogous to word2vec \cite{mikolov2013distributed}

\subsection{Context Nodes}
We generate context nodes for each individual source node by performing $L$ random walks of fixed walk length ($l$) \footnote{In our experiments we have set $l=40$ and $L=10$.} with the source node as the starting vertex.  Similar core nodes can sometimes be adjacent to the source node or they can be separated by multiple hops. Hence in our exploration strategy we assume roles of both forms of extreme sampling strategies -- the breath-first sampling as well as the depth-first sampling~\cite{grover2016node2vec}.

\section{Methodology}
\label{sec:method}

Consider a random surfer which has started from node $i$ and is currently at node $j$, where it is not necessary that  $(i,j) \in E$. The decision for the next destination $(k)$ for the random walk given that $(j,k) \in E$ is given by $p_{i,j,k} = \frac{\pi_{ijk}*w_{jk}}{Z}$ where $\pi_{ijk}$ denotes the unnormalized transition probability, $Z$ is the normalization constant and $w_{j,k}$ is the weight of edge $(j,k)$. $w_{j,k}$ is 1 in case the graph is unweighted. The unnormalized transition probability for our approach is given below. 

\[
\small
    \pi_{i,j,k}= 
\begin{cases}
    \frac{1}{(|c_i - c_j|+1)*\mathcal{D}*\lambda} & (j, k) \in E, i = k; \\
     1 & (j, k) \in E , i \neq k;  \\
     \frac{1}{(|c_i - c_j|+1)*\mathcal{D}*\gamma} & (j, k) \in E, i \neq k, (k, i) \notin E \\
    0,              & \text{otherwise}
\end{cases}
\]

Here $p_{i,j,k}$ is the probability of the random surfer starting from vertex $i$, currently at vertex $j$ to transition to vertex $k$. $c_j,c_k$ signifies the core id for vertex $j,k$ respectively. $\lambda$ and $\gamma$ can be tuned for the  purpose of exploring in the close neighborhood of the source node or traverse  distant neighborhood of the source node. The pseudocode for {\em core2vec} is presented in  Algorithm~\ref{Algo:core2vec}.

\begin{algorithm} 
\tiny
\caption{The core2vec algorithm}\label{Algo:core2vec}
\begin{algorithmic}
\Procedure{kCore}{}
\State Input: $\mathit {Graph}$, $G(V,E)$
\State Output: $C[K], K = |V|$
\State $k \gets 1$
\While {$|V| \geq 0 $}
\While {true}
\State remove all vertices with degree $ \le k$
\State until all remaining vertices have degree $\geq k$ 
\State $ \forall$ vertex ($v$) removed , $C[v] \leftarrow k$
\EndWhile
$k\leftarrow k+1$
\EndWhile
\State Return $(C)$
\EndProcedure
\end{algorithmic}
\hrulefill

\begin{algorithmic}

\Procedure{LearnFeatures} {$G = (V,E,W$), dimensions $d$, walks per node $L$, walk length $l$, context size $c$,  exploration parameters $\lambda$, $\gamma$, penalty parameter $\mathcal{D}$, probability transition matrix $\mathcal{P}$}
\State ${core\_dict} \gets KCORE(G)$
\State $\mathcal{P} \gets \text{PreprocessProb$(G,\lambda,\gamma,\mathcal{D},\mathit{core\_dict})$}$
\State $G' \leftarrow (V,E,\mathcal{P})$
\For{$\textbf{all } \textnormal{nodes } u \in V $}
\State Initialize $walks$ to \textit{empty}
\State $walks \leftarrow \text{genWalks$(G', u, l,L)$}$
\EndFor
\State $f \gets \text{SGD$(c, d, walks)$}$ 
\State \Return{$f$}
\EndProcedure
\\
\hrulefill

\end{algorithmic}

 \begin{algorithmic}
 \Procedure{genWalks}{}
 \State Input: $G'(V,E,\mathcal{P})$,  \textit{start node} $(\mathit{u})$, \textit{walk length} $(\mathit{l})$, \textit{total walk} $(\mathit{L})$
 \State Output: $\mathit{walk}$
 \State Initialise $\mathit{walk} \leftarrow {u}  $
 \State Intialise $\mathit{walks} \leftarrow \{\}$
 \For { $num\_walks = 1$ \textbf{to} $L$ }
 \For {$walk\_iter = 1$ \textbf{to} $l$ }
 \State $curr = walk[-1]$
 \State $\mathcal{N}_{curr} \leftarrow Neigbours\_set(curr,G)$
 \State$ s \leftarrow Sample(\mathcal{N}_{curr},\mathcal{P})$
 \State $walk \leftarrow s$
 \EndFor
 \State $walks \leftarrow walk$
 \EndFor
 \State \Return $\mathit{walks}$
 \EndProcedure
 \end{algorithmic}
 \end{algorithm}

\section{Datasets}
\label{sec:dataset}
\subsection{Network data used for experiments}
We use two very well-known network datasets -- \textbf{Les Miserables (Lemis)} and \textbf {Jazz musicians (Jazz)} as benchmarks to carry out our experiments to show the efficacy of our approach. Datasets have been taken from ~\cite{kunegis2013konect}.

\subsection{Dataset for validation}
\subsubsection{Training data}
We use the English word association data collected from the two notable crowd-sourcing efforts -- (i) University of South Florida word association project (USF) and (ii) small world of word project (SWOW)\footnote{http://www.smallworldofwords.com/new/visualize/}. In each case, a group of participants were given a cue word and asked to report the first few words that come to their mind in response to the cue. Normative forms of words were reported by each project. Details about the data can be obtained from ~\cite{nelson2004university,de2016predicting}.

\subsubsection{Test data}

We use three datasets from where we obtain the ground-truth similarity between word pairs. These are: (i) Mturk-771~\cite{halawi2012large}, (ii) WordSim353~\cite{finkelstein2001placing} and (iii) SimLex-999~\cite{hill2015simlex}. Mturk-771 and WordSim353 score words on both similarity as well as relatedness. However SimLex-999 scores words only on the basis of high semantic similarity. 

\section{Experiments}
\label{sec:experiments}

We establish the necessity of our approach based on the following two metrics -- (i) closeness and (ii) separability. 

\subsection{Closeness $(\mathcal{C})$}
This metric estimates the cohesiveness of a core's nodes around its \textit{centroid} (i.e., the mean of all the vectors corresponding to the nodes within a core). Closeness is calculated as the average of cosine similarities of all nodes in a core with that of the core's centroid. The higher the value of closeness the more compact the core is.

\subsection{Separability $(\mathcal{S})$}
Separability determines the overall separation among the different distinct cores. This is calculated as the average Euclidean distance between pairwise core centroids. The higher the value of separability the more well-separated the cores are.

 The different methods are compared in Table~\ref{Tab:Closeness}. The closeness and separability increases by tuning the penalty parameter $(\mathcal{D})$ and in both networks we obtain better scores compared to different naive random walk approaches like node2vec, DeepWalk and LINE.

\begin{table}[h!]
\caption{The values of $\mathcal{C, S}$ for different methods. {In case of core2vec, $\mathcal{D}=3.5$, $\lambda=0.35$ and $\gamma =2.5$.}}
\centering
\renewcommand{\arraystretch}{1.8}
\begin{tabular}{ | c || c | c | }
 \hline
 \hline
 Algorithm & Les Miserables network & Jazz musicians' network \\
 \hline
\cellcolor{white}core2vec  & \cellcolor{white}$\mathbf{0.124, 0.414}$  & \cellcolor{white}$\mathbf{0.543, 0.404}$\\

node2vec & 0.097, 0.357 & 0.422, 0.276\\

DeepWalk & 0.076, 0.311 & 0.317, 0.239\\

LINE& 0.085, 0.345 & 0.329, 0.255\\
 \hline
\end{tabular}
\label{Tab:Closeness}
\end{table}

\section{Validation}
\label{sec:eval}
\subsection{Word association networks}
Linguists and cognitive psychologists report~\cite{hills2013company,benedek2017semantic,de2016predicting} that by the time children are 4 years old they hear approximately 10 to 50 million words which only increases manifold as their age progresses. The ability of humans to learn and recall such massive information is through associations. This claim is supported by the finding that word pairs that are semantically associated but mean different things, such as ``sky", ``blue" or ``banana", ``yellow"  activate the same regions of the human brain. Networks constructed from sentimentally aligned words which have similar associations or relatedness, have dense cores of highly connected words also known as kernel lexicons~\cite{choudhury2010global,i2001small} linked with a relatively sparse periphery.

\subsection{Outline of the validation framework}

We hypothesize that for word association networks with well-defined core-periphery structures, the embeddings obtained from core2vec should be more representative than the state-of-the-art methods like node2vec, DeepWalk and LINE.

In order to establish the above hypothesis we first obtain embeddings of nodes for each of the two word association networks introduced in section~\ref{sec:dataset}. We obtain the embeddings using core2vec as well as the other baselines -- core2vec, DeepWalk and LINE. 

Next we consider three ground-truth datasets -- SimLex-999, WordSim-353 and Mturk-771 that contain a set of word pairs and their similarity/relatedness scores. We rank these word pairs based on these scores. In parallel, we obtain the similarities of exactly these word pairs by estimating the cosine similarities of their corresponding embeddings obtained from the word association networks. We again rank these word pairs based on these cosine similarities. Finally, we estimate the Spearman's rank correlation coefficient between the two rankings (one from the ground-truth similarities and the other from the embedding similarities). 

\subsection{Results}

The key results for the two different association networks - SWOW and USF -- are shown in Tables~\ref{Tab:small} and~\ref{Tab:large} respectively. The correlation values in the tables indicate that core2vec outperforms all the baselines. The $p$ values are further noted to demonstrate that our observations are significant. Depending from where the ground-truth similarities are drawn, in some cases we even obtain an improvement as high as \textbf{46\%}. An interesting point is that the best benefit of core2vec is obtained when the ground-truth similarities are drawn from the SimLex dataset. This shows that core2vec is able to better capture strong semantic similarities in comparison to mere relatedness. 

\begin{figure}[h!]
  \centering
  	\includegraphics[width=0.38\textwidth,height=4.5cm]{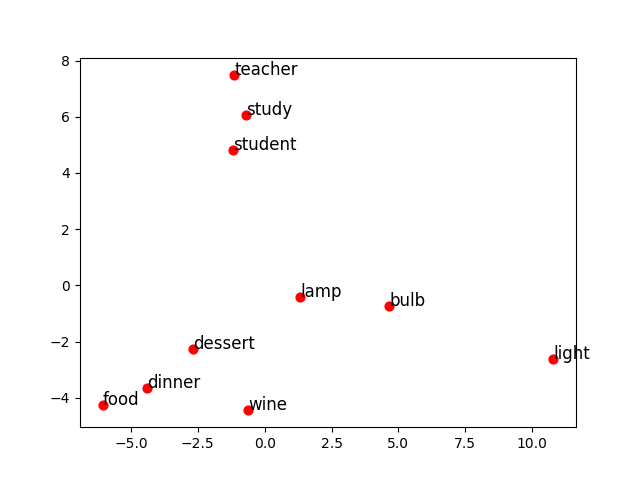}
     \includegraphics[width=0.38\textwidth,height=4.5cm]{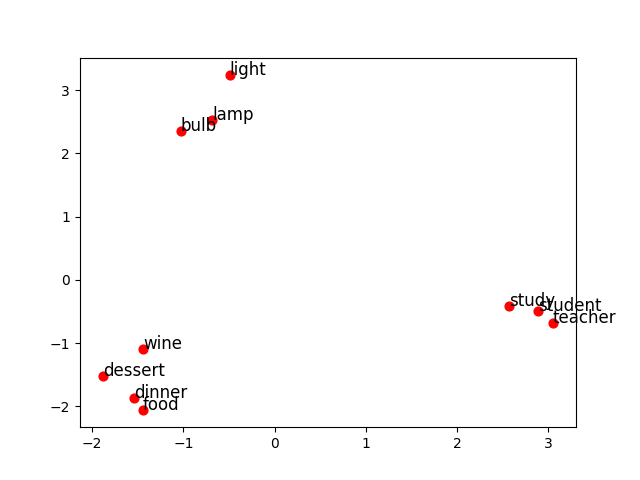}
    \caption{\label{Fig:word_plot}core2vec brings semantically similar words closer in the vector space. {The values of $\mathcal{D}$, $\lambda$ and $\gamma$ are 3.5, 0.3 and 3 respectively.}}
\end{figure}

To further understand our results, we plot in Figure~\ref{Fig:word_plot}, the embeddings learned by {\em node2vec} (top panel) and {\em core2vec} (bottom panel) for the same set of words projected on a 2D plane (using PCA). Figure ~\ref{Fig:word_plot} clearly shows that words with similar meanings or words which usually have more similar contexts are noticeably clustered better in case of {\em core2vec}.

\begin{table}[h!]
\footnotesize
\caption{Results (Spearman's correlation coefficient, $p$ value of significance) for SWOW word association network. {The values of $\mathcal{D}$, $\lambda$ and $\gamma$ are 3.5, 0.3 and 3 respectively.}}
\centering
\renewcommand{\arraystretch}{1.8}
\begin{tabular}{ | p{1.1cm} || p{2.05cm} | p{2.05cm} | p{2.175cm} |  }
 \hline
 \hline
 Algorithm & SimLex & WordSim353 & Amazon MTurk\\
 \hline
\cellcolor{white}core2vec  & \cellcolor{white}$\mathbf{0.548, 3.34 e^{-79}}$  & \cellcolor{white}$\mathbf{0.654, 2.49 e^{-42}}$ & \cellcolor{white}$\mathbf{0.692, 1.33 e^{-109}}$\\

node2vec & 0.467, $2.00 e^{-55}$ & 0.640, $3.72 e^{-40}$  & 0.664, $3.05 e^{-98}$\\

DeepWalk & 0.444, $9.80 e^{-50}$ & 0.639, $5.86 e^{-40}$&  0.653, $3.49 e^{-94}$\\

LINE& 0.449, $9.53 e^{-51}$ &0.635, $2.14 e^{-39}$ & 0.571, $3.01 e^{-67}$ \\
 \hline
\end{tabular}
\label{Tab:small}
\end{table}

\begin{table}[h!]
\footnotesize
\caption{Results (Spearman's correlation coefficient, $p$ value of significance) for USF word association network. {The values of $\mathcal{D}$, $\lambda$ and $\gamma$ are 3.5, 0.3 and 3 respectively.}}
\centering
\renewcommand{\arraystretch}{1.8}
\begin{tabular}{ | p{1.1cm} || p{2.05cm} | p{2.05cm} | p{2.175cm} |  }
 \hline
 \hline
 Algorithm & SimLex & WordSim353 & Amazon MTurk\\
 \hline
\cellcolor{white}core2vec  & \cellcolor{white}$\mathbf{0.425, 2.11 e^{-39}}$  & \cellcolor{white}$\mathbf{0.476, 5.29 e^{-32}}$ & \cellcolor{white}$\mathbf{0.621, 1.33 e^{-52}}$\\

node2vec & 0.136, $2.00 e^{-15}$ & 0.439, $4.52 e^{-30}$  & 0.593, $7.35 e^{-48}$\\

DeepWalk & 0.116, $9.80 e^{-20}$ & 0.429, $2.66 e^{-25}$&  0.604, $1.79 e^{-48}$\\

LINE& 0.111, $9.53 e^{-12}$ &0.424, $5.94 e^{-24}$ & 0.598, $9.26 e^{-47}$ \\
 \hline
\end{tabular}
\label{Tab:large}
\end{table}

\section{Discussion}
\label{sec:diss}
\subsection{Hyper-parameters}
Our framework has three hyper-parameters -- $\lambda,\gamma$ and $\mathcal{D}$. Low values of $\lambda$ will restrict exploration strategy  preferentially within close proximity. Low values of $\gamma$ will result the random walk sample neighbors from distant hops. The penalty parameter $\mathcal{D}$ penalizes random runs of high core difference. After extensive experimentation we observe that the hyper-parameter values that work best for a network is dependent on the structure of the network being considered. However, increasing $\mathcal{D}$ does not indefinitely increase closeness and separability. A systematic grid search allows us to identify the best choice for each network. 

Further, note that  higher values of the walk length ($l$) and number of walks ($L$) usually yield better results. This is very similar to other exploration based embedding techniques~\cite{perozzi2014deepwalk,grover2016node2vec}.

\subsection{Scaling experiments}
Here we attempt to empirically test how well our model scales with respect to the number of nodes in the graph. We note that the theoretical time complexity of the KCORE algorithm is linear in the number of edges (i.e., $O|E|$) which would make the overall complexity of our model at least $O|V^2|$. However, since most large graphs under study are sparse our algorithm completes in sub-quadratic time in practice.

We consider $6$ Erdos-Renyi random networks. {Probability of edge formation is set to $\frac{\hat{k}}{N}$ where $\hat{k}$ is the average degree and is set to $30$. $N$ is the number of nodes which varies assuming values like 10, 100, 300, 1000, 3000, 30000. Note that the logarithms of these numbers increase from 1 to 4.5 linearly. We record the logarithm of the time taken by our algorithm to run with these networks as inputs and plot the result in Figure~\ref{fig:scaling}. The plot is close to linear with slope one indicating that the time taken by core2vec is mostly linear in the number of nodes.

\begin{figure}[h!]
  \centering
     \includegraphics[width=0.38\textwidth,height=3cm]{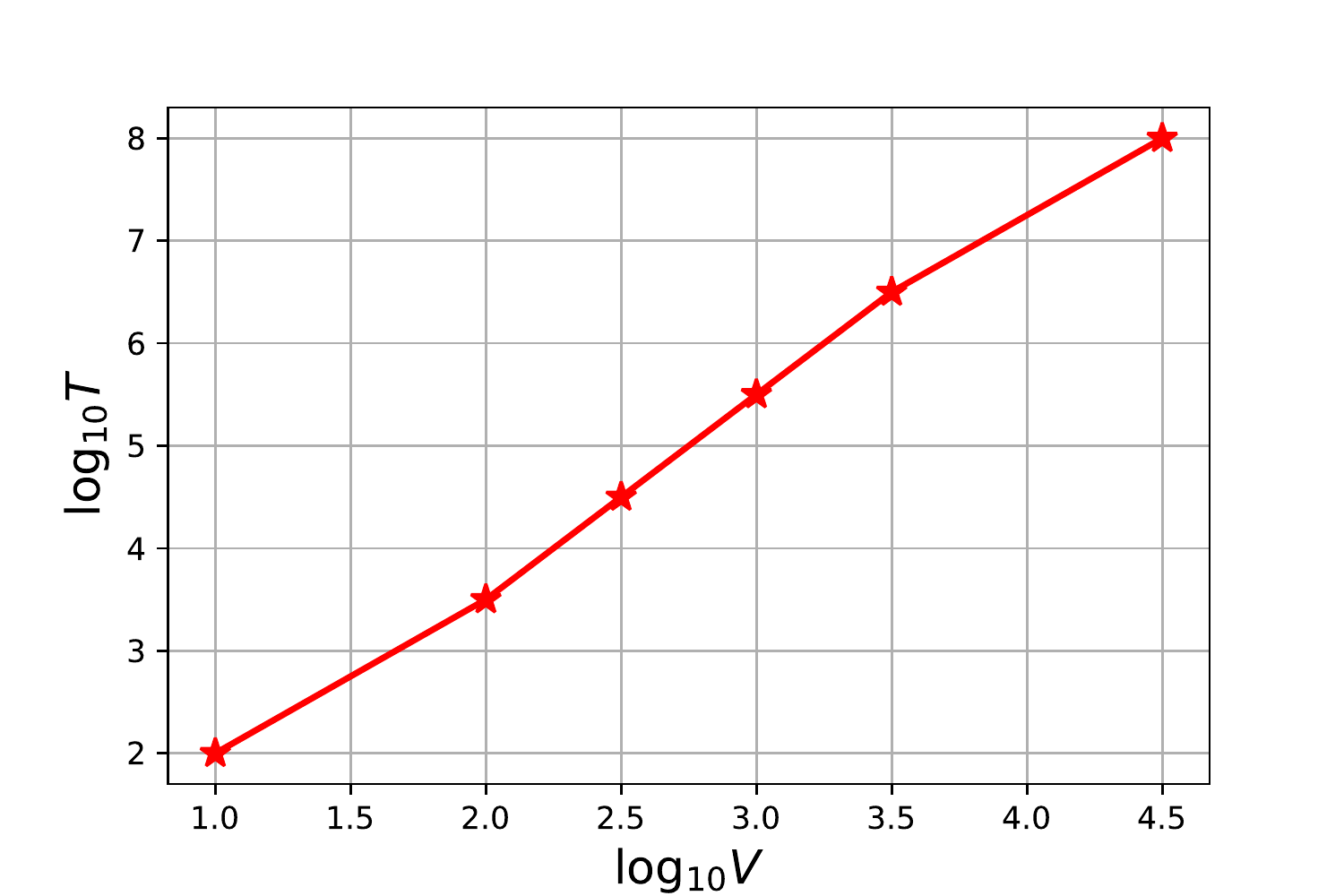}
    \caption{\label{fig:scaling} Logarithm of the time taken by core2vec vs $log_{10}|V|$.}
\end{figure}

\section{Conclusion}

This paper to the best of our knowledge is a first work which demonstrates a network embedding task, utilizing global information like coreness of a node. We successfully show that our embedding approach brings similar core nodes together in latent dimensions and separates disparate core nodes. 

We apply our embedding approach to the task of detecting similar words by training on two large word association networks. Embeddings obtained by our approach maps similar words closer in space compared to other baseline approaches.    
\bibliographystyle{abbrv}

\bibliography{Trans}

\end{document}